\documentclass[%
					 aip,%
					 jcp,%
					 amsmath,%
					 amssymb,%
					 reprint,%
					 author-year,%
					 author-numerical,%
					 ]{revtex4-1}  

\usepackage{graphicx}
\usepackage{dcolumn}
\usepackage{bm}
\usepackage{url}
\usepackage{epsfig}

\begin{document}
\preprint{AIP/123-QED}
\title[\textit{A. Metere, et al.} - Formation of a columnar liquid crystal in simple monatomic system of particles]{Formation of a columnar liquid crystal in a simple one-component system of particles}

\author{Alfredo Metere}
\email{alfredo.metere@mmk.su.se}
\affiliation {Department of Materials and Environmental Chemistry, Stockholm University, S-106 91, Stockholm, Sweden}

\author{Sten Sarman}
\affiliation {Department of Materials and Environmental Chemistry, Stockholm University, S-106 91, Stockholm, Sweden}

\author{Tomas Oppelstrup}
\affiliation {Lawrence Livermore National Laboratory - 7000 East Avenue, Livermore, California 94551, USA}

\author{Mikhail Dzugutov}
\affiliation{Department of Mathematics and Centre for Parallel Computers, Royal Institute of Technology, S-100 44 Stockholm, Sweden}


\begin{abstract}
  We report a molecular dynamics simulation demonstrating that a
  columnar liquid crystal, commonly formed by disc-shaped molecules, can
  be formed by identical particles interacting via a spherically
  symmetric potential. Upon isochoric cooling from a low-density
  isotropic liquid state the simulated system performed a weak first
  order phase transition which produced a liquid crystal phase
  composed of parallel particle columns arranged in a hexagonal
  pattern in the plane perpendicular to the column axis. The particles
  within columns formed a liquid structure and demonstrated a
  significant intracolumn diffusion. Further cooling resulted in
  another first-order transition whereby the column structure became
  periodically ordered in three dimensions transforming the
  liquid-crystal phase into a crystal. This result is the first
  observation of a liquid crystal formation in a simple one-component
  system of particles. Its conceptual significance is in that it
  demonstrated that liquid crystals that have so far only been
  produced in systems of anisometric molecules, can also be formed by
  mesoscopic soft-matter and colloidal systems of spherical particles
  with appropriately tuned interatomic potential.
  \end{abstract}

\date{\today}
\maketitle

\section{Introduction} 
Liquid crystals \cite{degennes, chandrasekhar} are anisotropic phases
which combine fluidity with periodicity in less than three
dimensions. One-dimensional periodic order in the smectic phases
arises from uniaxial stacking of liquid layers. Its spatial extent,
however, is limited due to Landau-Peierls instability \cite{LAN},
whereas the two-dimensional periodic order characteristic of columnar
liquid crystals \cite{chandrasekhar, BOCK} is stable at the global
scale. The experimentally observed columnar liquid crystals represent
close packing of parallel columns composed of axially stacked
molecular units. In the plane perpendicular to the column axis, the
column packing forms a regular pattern with two-dimensional
periodicity, whereas the remaining continuous translational symmetry
dimension is directed along the column axis. The columns are commonly
formed by disk-like molecules \cite{BOCK} or wedge-shaped dendrons
\cite{ZENG}. The close packing of columns fixes the position of a
molecule in the plane perpendicular to the column axis, but the
molecules' stacking along the axis is irregular, and the axial
position of a molecule is not defined with respect to its neighbors in
adjacent columns, which gives rise to the continuous translational
symmetry in the axial dimension.

Particle simulations have proved to be an indispensable tool for
understanding the relationship between the phase behaviour and the
molecular-level properties of liquid crystals \cite{care, bates, zannoni}. These simulation have so far followed the phenomenological
paradigm that dominated the science of liquid crystals for decades,
whereby the structural anisotropy of liquid crystals was assumed to be
determined by anisometric shape of a molecule. Respectively, the
shapes of the constituent particles in the models of liquid crystals
have been designed to imitate the shapes of the molecules in the
respective mesogens: the smectic phases have been simulated using
rod-like particles \cite{bolhuis,Allen, sarman, sarman2}, and models of
columnar liquid crystals \cite{zannoni, MARTIN, bates2, ANDRI, veerman}
commonly used flat discotic particles or oblate ellipsoids.

That approach is based on the conjecture according to which formation
of the structurally anisotropic liquid crystals is driven by the
entropic component of the free energy \cite{LUB}. The origin of this
line of thinking can be traced to the seminal work of Onsager
\cite{ONS} which linked the anisotropy of the equilibrium structure in
a liquid of rods to the entropy of packing.  The conjecture of the
entropic mechanism of liquid-crystal formation have been further
strengthen by simulations using hard anisometric particles \cite{
  Allen, veerman}  stressing the role of the geometry of excluded
volume.

A question of general conceptual interest for the statistical
mechanics of condensed matter is whether the anisometry of the
mesogenic molecules is really a prerequisite for producing structural
anisotropy in liquid phases, as it is conjectured by the entropic
paradigm, or the entropic effects on the structure formation due to
the particle geometry can be compensated by an appropriately designed
spherically symmetric interaction potential. This question is confined
to the positional structural ordering; the nematic orientational ordering in liquids of anisometric particles is obviously beyond the scope of present discussion.

This question has been addressed in a molecular-dynamics simulation
that we report here. It is demonstrated that a single-component system
of particles interacting via a spherically-symmetric potential can
form a a thermodynamically equilibrium hexagonal columnar liquid
crystal. This mesophase was formed a result of a first-order phase
transition which was observed when the system had beed cooled
isochorically at low density from its equilibrium isotropic liquid
state. A distinctive feature of this columnar phase is the liquid
structure of its columns and considerable particle diffusion along the
axis. Under further cooling, another phase transition took place,
transforming the columnar liquid crystal into a columnar structure
with global three-dimensional periodicity. This result is the first
compelling evidence that an anisotropic liquid crystal phase can be
produced in a system of identical particles interacting via a
spherically symmetric potential. This finding opens a possibility of
producing similar columnar liquid-crystal phases in colloidal and
soft-matter systems composed of spherically-shape particles.

\section{Model and simulation}

\begin{figure} 
\centering
\includegraphics[width=8cm]{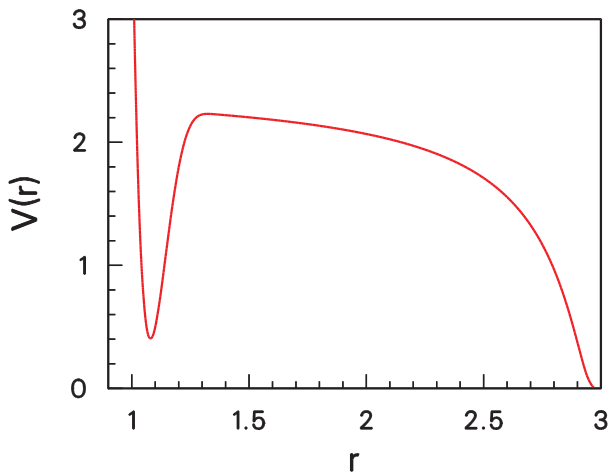}
\caption{Pair potential}
\label{fig1}
\end{figure}

We report here a molecular-dynamics simulation of a simple
single-component system of particles interacting via the pair
potential shown in Fig. \ref{fig1}. The functional form of the
potential energy for two particles separated by the distance $r$ is
defined as:
\begin{equation} 
V(r)  = a_1  ( r^{-m} - d) H(r,b_1,c_1) + a_2 H(r,b_2,c_2),  \\ 
\end{equation}
where
\begin{equation} 
H(r,b,c) = \left\{ \begin{array}{ll}
  \exp\left( \frac{b}{r-c} \right) & r < c,\\
0 & r \geq c.
\end{array}
\right.       
\end{equation} \\

The values of the parameters are presented in Table \ref{table1}. The first
term of this functional form describes the short-range repulsion
part of the potential, and its first minimum, whereas the second
term is responsible for the long-range repulsion.  All the
quantities we report here are expressed in the reduced units used
in the definition of the potential. We also note that the
short-range repulsion part of the potential, and the position of
its first minimum closely approximate those in the Lennard-Jones
(LJ) potential \cite{hansen}, which makes it possible to directly
compare the reduced number densities, and other thermodynamic
parameters of the two systems. 

It has to be mentioned that this pair potential represents a
modification of an earlier reported potential \cite{ME} which was
found to produce the smectic-$B$ crystal. The main difference between
the two potentials is that in the present one the long-range repulsive
part extends to a significantly larger distance. By increasing the
long-range separation of the interacting particles, this modification
of the potential was intended to reduce the density of particle
packing in low-temperature phases where that part or interaction
energy becomes significant.

\begin{table}
\begin{tabular}{cccccccc}
 \hline 
\hline 
m & $a_1$  & $b_1$ & $c_1$ & $a_2$ & $b_2$ & $c_2$ & $d$ \\
\hline 
\hline 
12 \space \space & 265.85 \space \space & 1.5 \space \space & 1.45 \space \space & 2.5 \space \space &
0.19 \space \space & 3.0 \space \space & 0.8 \\
\hline 
\end{tabular}
\caption{Values of the  parameters for the pair potential.} 
\label{table1}
\end{table}

The described pair potential was exploited in a molecular dynamics
model comprised of 16384 identical particles confined to a cubic box
with periodic boundary condition. In this simulation, we explored the
phase transformations of the described system by changing its
temperature isochorically at the reduced number density
$\rho=0.3$. Note that this density is very low as compared with the
density of the LJ liquid at its triple-point, $\rho=0.84$
\cite{hansen}. The temperature was changed in a stepwise manner; each
temperature step was followed by a comprehensive equilibration run
that typically amounted to $10^7-10^8$ simulation timesteps. The
temperature adjustments were performed by appropriately scaling the
particle velocities.

\section{Results}

\begin{figure} 
\includegraphics[width=8cm]{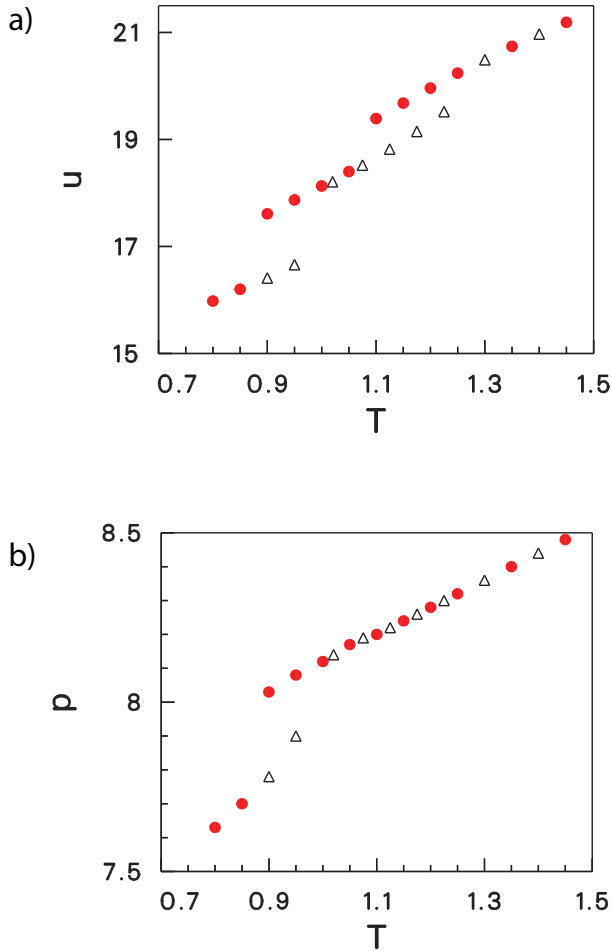}
\caption{ Isochoric liquid-solid phase transformation. $a)$ and $b)$,
  respectively, depict energy and pressure variation as a function of
  temperature. Dots and open triangles correspond to cooling and
  heating, respectively.}
\label{fig2}
\end{figure}

We began the simulation by equilibrating the system in its stable
isotropic liquid state at sufficiently high temperature at the number
density $\rho=0.3$, which was followed by an isochoric cooling
performed according to the described step-wise procedure. A
discontinuous reduction of energy characteristic of the first-order
phase transition was detected when the system was cooled below
$T=1.1$, Fig. \ref{fig2}. This thermodynamic singularity was
accompanied by a sharp reduction of the diffusion rate. The
first-order nature of the observed transition was further confirmed by
a significant hysteresis which was detected when reheating the
low-temperature phase, Fig. \ref{fig2}. An interesting peculiarity of this
phase transition is that no discernable singularity in the pressure variation have
been found within that range of temperature. This possibly indicates a
weak nature of this first-order transition \cite{muratov}. The
possibility of this kind of phase behaviour has been theoretically
conjectured for columnar liquid crystals \cite{ grason}.

\begin{figure} 
\centerline{\includegraphics[width=7.cm]{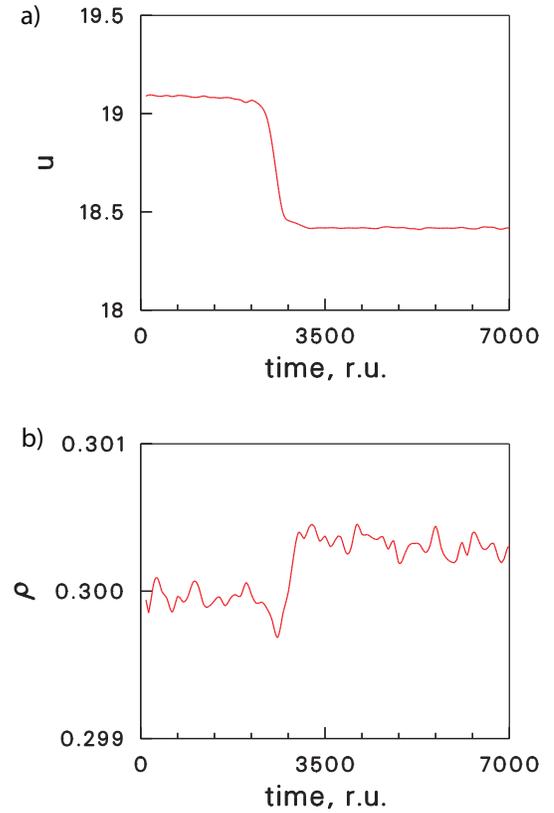}}
\caption{Transformation of the isotropic liquid into columnar liquid crystal at constant pressure $P=8.16$ and temperature $T=1.05$. $a)$ and $b)$ plots, respectively, depict variations of energy and density}
\label{fig3}
\end{figure}

We have also explored that phase transformation at constant pressure $P=8.16$ at the transition temperature $T=1.05$, see Fig. \ref{fig3}. The pressure was fixed using the technique suggested by Berendsen et al \cite{BERENDSEN}. The extremely small value of the density step upon the transition, about $0.1 \%$, is consistent with the isochoric results in Fig. \ref{fig2}.

Further cooling of this phase resulted in another first-order
phase transition; in this case, a clear hysteresis in both energy
and pressure was observed, see Fig. \ref{fig2}. This transition
further reduced the diffusion rate to the value characteristic of
a solid phase.  Thus produced low-temperature phases will
hereinafter be referred to as Phase I and Phase II, according to
the order of their occurrence upon cooling. Evidently, the Phase
I remains in a thermodynamically stable equilibrium within a
finite range of temperature.

The structure characterisation of the two phases has first been
performed by analysing the pattern of their density correlations
in the the Fourier-space. For that purpose, we calculated the
structure factor $S({\bf Q}) = \langle \rho({\bf Q})\rho(-{\bf
  Q})\rangle$, where $\rho({\bf Q})$ is a Fourier-component of
the  number density of a system of $N$ particles:
\begin{equation}
\rho({\bf Q}) = \frac{1}{\sqrt{N}} \sum_{j=1}^{N}\exp (-i{\bf Q r}_j), 
\end{equation} 
${\bf r}_i$ being the positions of the system's particles, and
$\langle \rangle$ denote ensemble averaging. $S({\bf Q})$
represents the diffraction intensity as measured in diffraction
experiments.

\begin{figure}  
\centerline{\includegraphics[width=8cm]{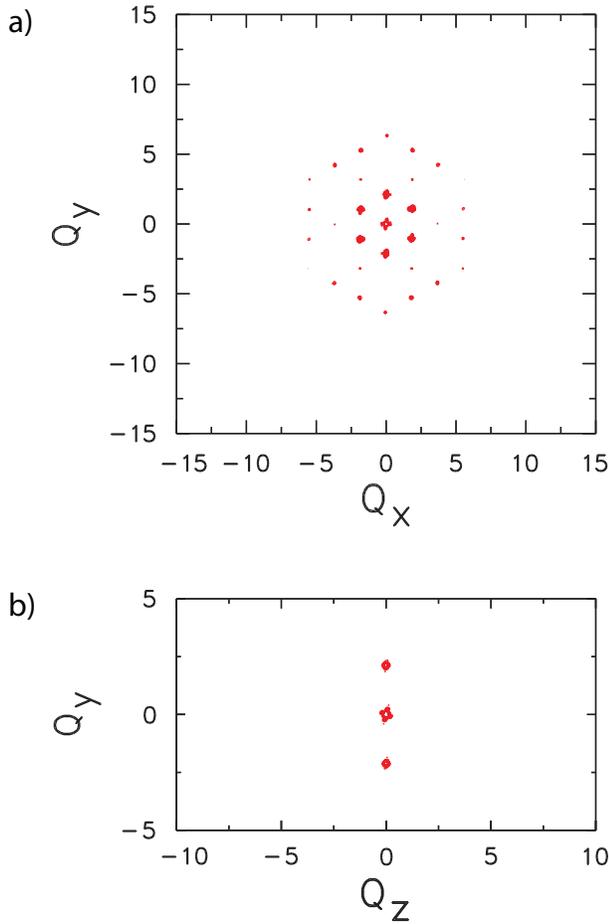}}
\caption{ The isointensity plots of the structure factor $S(\bf Q)$
  calculated for a Phase I configuration in two orthogonal
  reciprocal-space planes. From top to bottom, respectively: $Q_z=0$
  and $Q_y=0$.  $Q_z$ denotes the axial dimension, and $Q_y$
  corresponds to one of the translational symmetry vectors orthogonal
  to the axis.}
\label{fig4}
\end{figure} 

The structure factor was first calculated on the $\bf Q$-space sphere
of the radius corresponding to position of the first peak of the
spherically averaged $S({\bf Q})$. A well-defined pattern of
$S({\bf Q})$ maxima was observed, which made it possible to
determine the global symmetry and the axis. The axis orientation
having been found, $S({\bf Q})$ was calculated, for each phase,
in two characteristic $Q$-planes: $Q_z=0$, $Q_z$ being the axis
coordinate, and $Q_y=0$, $Q_y$ coordinate corresponding to a
translational symmetry vector orthogonal to the axis. The
results, for both phases, are shown in Fig. \ref{fig4}.

\begin{figure}  
\includegraphics[width=8.cm]{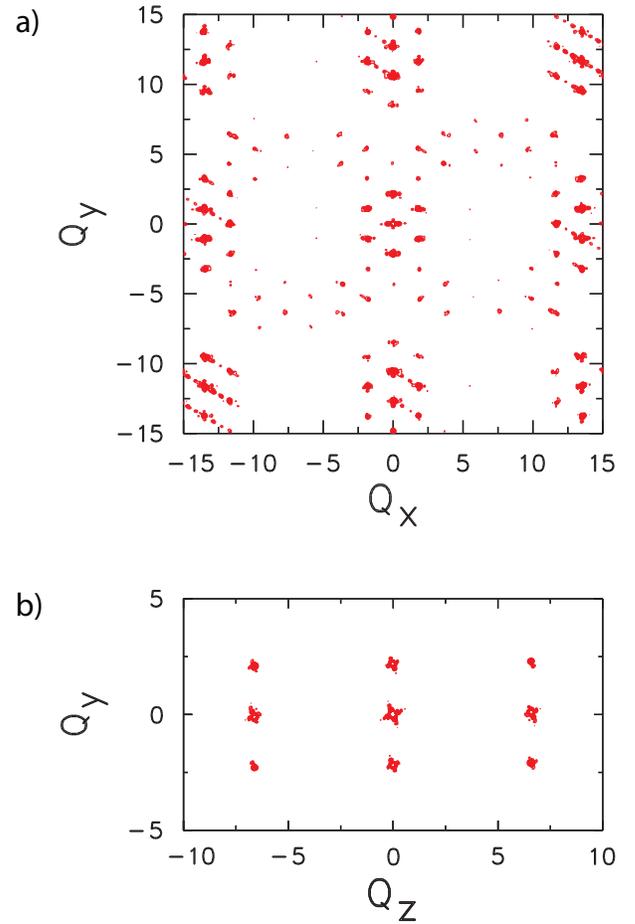}
\caption{The isointensity plots of the structure factor $S(\bf Q)$
  calculated for a Phase II configuration in two orthogonal
  reciprocal-space planes. $a)$ and $b)$, respectively: $Q_z=0$
  and $Q_y=0$.  $Q_z$ denotes the axial dimension, and $Q_y$
  corresponds to one of the translational symmetry vectors orthogonal
  to the axis.}
\label{fig5}
\end{figure} 

These diffraction results lead us to making the following
conclusions. First, the Phase I exhibits two-dimensional global
hexagonal periodicity in the plain perpendicular to the axis, whereas
no structure has been found in the axial dimension. Thus, Phase I
appear to be a columnar liquid crystal where parallel columns form a
hexagonal close-packed pattern in the plane perpendicular to the
column axis. In the axial dimension, the phase remain structureless.
The transformation of Phase I into Phase II upon further cooling
breaks the continuous translational symmetry in the axial
dimension. Moreover, the crystalline order within column structure is
induced in all three dimensions. This ordering within columns
apparently occurs in a globally coherent manner, producing both axial
periodicity and a periodic pattern in the plane perpendicular to the
axis. The latter too is arising due to the coherence of regular
packing of particles within column in Phase II. This global order is
indicated by additional sets of diffraction peaks which appear in both
$\bf Q$-planes. Thus, Phase II appear to be a true crystal,
composed of coherently arranged crystalline columns organised in the
same hexagonal pattern as in Phase I. We also note that in both phases
the intercolumn separation as inferred from the diffraction pattern is
consistent with the long-range repulsion range of the pair potential,
Fig. \ref{fig1}.

\begin{figure}  
\centering
\includegraphics[width=8cm]{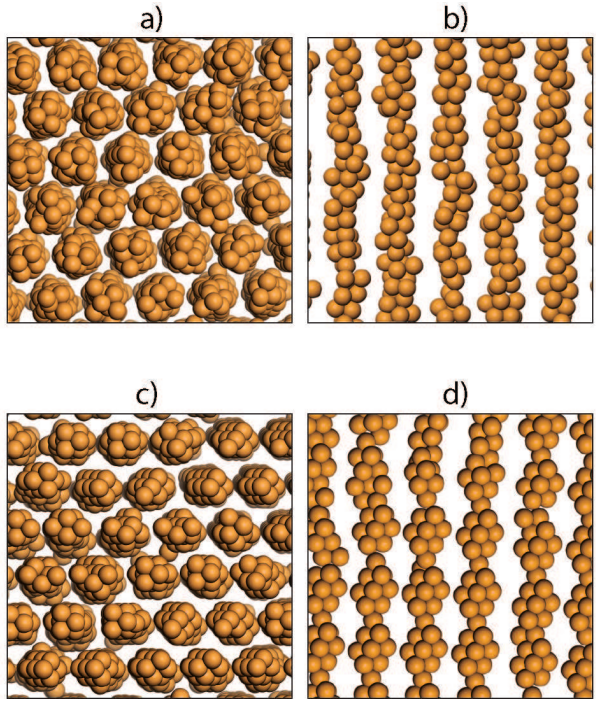}
\caption{ The real-space images of the two phases
  configurations. a) and c): view from the axial direction, Phase
  I and Phase II, respectively. b) and d): a one-column layer of
  the structure, cut parallel to the layer axis, as viewed
  perpendicular to the axis. The particle diameter in the plot is
  assumed to be $1$, in reduced units, which corresponds to the
  particles separation distance at the hard-core contact, see the
  pair potential in Fig.1}
\label{fig6}
\end{figure}

The conclusions about the structure of the two phases  inferred
from the diffraction data are consistent with the real-space images of
the configurations of these phases which are shown in
Fig.  \ref{fig6}. Both Phase I and Phase II appear to be columnar
structures where parallel particle columns are arranged in
triangular-hexagonal pattern in the plane perpendicular to the column
axis.

The most remarkable structural aspect of the Phase I that can be
observed in its images in Fig. \ref{fig6} is that the configuration of a column
can be described as three-dimensional liquid-like dense random
particle packing. This column structure is distinctively different
from that ubiquitously occurring in discotic liquid crystals where the
columns are composed of axially stacked disc-like particles. The
discotic columns are of one-particle width, and they are densely
packed laterally, forming a periodic pattern in the plane
perpendicular to the axis.  Under these constraints, the position of a
constituent particle in a discotic phase is limited both axially and
laterally. Such a structure can only allow solid-like vibrational
particle motions, and the continuous global translational symmetry of
the discotic phases in the axial dimension arises as a result of the
lack of coherence between columns in the stacking order.  By contrast,
the configuration of the Phase I shown in Fig. \ref{fig6} exhibits a
significant spacing between its columns, and each column appears to be
sufficiently wide to accomodate three-dimensional liquid-like dense
random particle packing.

\begin{figure}  
\centering
\includegraphics[width=8cm]{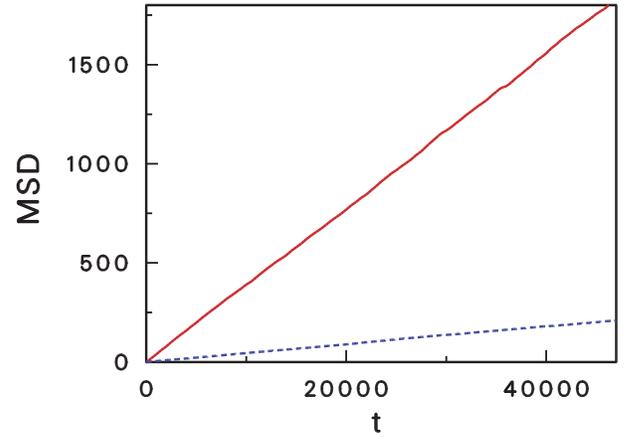}
\caption{ Mean-square particle displacement (MSD) in Phase I as a
  function of time at $T = 1$. Solid line corresponds to the diffusion
  along the column axis. Dashed line: MSD averaged over directions
  perpendicular to the axis}
\label{fig7}
\end{figure}
 
The characteristically liquid structure of the columns of Phase I
suggests that this phase should also be expected to sustain a kind of
liquid-like dynamics. Indeed, a considerable rate of intracolumn
liquid-like structural relaxation has been observed (these dynamics
can be seen in the movies included as supplementary material\cite{movie1}). It is
well known that atomic diffusion in dense fluids is driven by
collective local particle rearrangements and, in this way coupled to
the structural relaxation dynamics \cite{COH, ENT}. Driven by this
diffusion mechanism, a particle is expected to be able to diffuse to
unlimited distance within a column along the axis, whereas its
transition perpendicular to the axis is limited by the column
diameter. A particle can, nevertheless, leave the column by hopping to
an adjacent one; these hopping events, however, are expected to be
rare relative to the particle intra-column diffusive movements because
of the large energy cost of crossing the potential-energy maximum
separating the columns, see Fig.\ref{fig1} (the occasional intercolumn
particle hoppings can be observed in the movie included as
supplementary material). Driven by these two distinctly different
diffusion mechanisms, the particle diffusion in Phase I is therefore
expected to be strongly anisotropic, with its gradient directed along
the axis. This conjecture is confirmed by the mean-square particle
displacement data presented in Fig. \ref{fig7} which demonstrate that
particles indeed diffuse in the axial dimension much faster than in a
direction perpendicular to the axis.  These results convincingly
demonstrate that a constituent column in the Phase I represents a
genuine three-dimensional liquid, both structurally and dynamically.

One can also see in Fig. \ref{fig6} that the transition of the liquid
crystal (Phase I) into a crystalline phase possessing a
three-dimensional periodicity (Phase II) represents a structural
transformation of the liquid configuration of each column into a
three-dimensional periodic crystal configuration. It is also possible
to see that the axial periodic order of a crystalline column is
coherent with that of the neighbour columns, thereby producing a
global axial periodicity. In the reciprocal space pattern, this gives
rise to the respective diffraction peaks emerging in the axial
reciprocal-space plane, see Fig. \ref{fig5}. Moreover, the periodic order
arising within a column in the process of its crystallisation in the
directions perpendicular to the axis develops in a manner coherent
with the respective order in the neighbour columns. Thereby a pattern
of global periodicity with wave vectors defined by the close-neighbour
distance arises perpendicular to the axis, giving rise to a set of
additional diffraction peaks which appears in the respective
reciprocal-space plane as a result of the Phase II formation, see
Fig. \ref{fig5}.  We note that the planes of periodically stacked particle layers in Phase II that can be observed from the view perpendicular to the columns, Fig \ref{fig6}. A detailed analysis of this crystal structure will be reported elsewhere.

\section{Discussion}

Following the seminal work of Onsager \cite{ONS}, the science of
liquid crystals has been dominated by the entropic paradigm according
to which anisometric shape of the constituent particles is a
prerequisite for the formation of anisotropic structure in liquids
\cite{LUB}.  A conceptually significant implication of the present
result is that it directly contradicts that long-standing opinion. We
have demonstrated that an anisotropic pattern with two-dimensional
global periodicity can emerge in a liquid composed of a single sort of
particles interacting via a spherically-symmetric potential. Moreover,
this liquid phase has also demonstrated a strongly anisotropic
self-diffusion. This result thus makes it possible to conclude that
emergence of structural and dynamic anisotropy in a liquid phase is
not necessarily related to the entropic component in the free energy
that has so far been assumed to be responsible for the formation of
equilibrium anisotropic liquid phases.

Moreover, the Phase I represents a novel type of columnar liquid
crystal distinctively different from those commonly observed in
discotic systems. In the discotic columnar phases the time-averaged
positions of constituent particles are fixed due to the structural
constraints imposed by the intracolumn stacking order and the
close-packing of the columns. In contrast, the liquid crystal that we
discovered in this study is a genuine liquid phase, both structurally
an dynamically. Its columns demonstrate a three-dimensional
liquid-like structure and a high rate of structural relaxation
dynamics (see the movie in supplementary material) which mediates a
predominantly axial diffusion, see Fig \ref{fig7}. The possibility of
a liquid phase with uniaxial self-diffusion may be of a significant
technological interest.

Another interesting aspect of the liquid-crystal formation we report
here is its apparent similarity to the microphase separation
transition \cite{LEI}. The latter is commonly observed in some polymer
blends which, instead of macroscopic phase separation, are able to
form under cooling an equilibrium structure of microscopic-size
domains with different polymer concentration. These microdomains may
form different kinds of globally ordered superstructures, including
hexagonal symmetry similar to the one we observed in this
simulation. Both the structures produced by microphase separation and
the present columnar phase exhibit an anomalously low amplitude of
density modulation characteristic of weak crystallisation transition
\cite{muratov}. The simulation we report here thus demonstrates that
the microphase separation can be produced in a simple system of
identical particles using a potential with short-range attraction and
long-range repulsion.

One possible way of realisation of this model in terms of physical
systems is straightforward. Columnar liquid crystals have so far been
produced in colloidal systems of anisometric particles, commonly of
disk-like shape.  The present result suggests that columnar liquid
crystals with the structure similar to that we report here, and
possibly other liquid-crystal phases, can be formed by spherical
colloidal particles, with appropriate tuning of the effective force
field. It has to be mentioned that the main features of the pair
potential we exploited in this simulation are consistent with the
classical theory for colloidal interactions by Deryagin, Landau,
Verwey and Overbeek (DLVO) \cite{DL,DL2,DL3,MAL}, amended with hard-core
repulsion or steric repulsion close to the contact. We also mention
that a family of similarly-shaped pair potentials have been used to
produce patterns in two-dimensional simulations of soft-matter systems
\cite{RON}

\section{Conclusion}

In summary, we have presented a molecular dynamics simulation which
demonstrates that a thermodynamically equilibrium hexagonal columnar
liquid crystal can be formed in a simple single-component system of
particles as a result of a first-order phase transition from an
isotropic liquid phase. This result is the first demonstration that
liquid crystals that have so far only been found in mesogenes composed
of anisometric molecules can also be formed in a system of identical
particles with spherically-symmetric interparticle pair potential. Another
conceptually significant distinction of this liquid crystal as
compared with commonly observed discotic phases is that its columns
are structurally and dynamically fluid; we observed a considerable
diffusion along the column axis. The scope of impact of this finding
is expected to include a possibility of producing columnar liquid
crystals in colloidal systems of identical spherical particles and
other similar soft-matter mesoscopic particle systems.

\section{Acknowledgements} 

We thank Dr. B. Sadigh for very useful discussions.  This study
was supported by the Swedish E-Science Research Foundation
(SERC). Funding from the Swedish National Research Council (VR)
is gratefully acknowledged. This work has been approved for
release under Lawrence Livermore Release No. LLNL-JRNL-656140.

\end{document}